\documentclass[11pt]{article}
\usepackage{amsmath, slashed}
\usepackage{graphicx}
\usepackage{geometry}
\makeatletter
\g@addto@macro\bfseries{\boldmath} % use boldmath in titles
\makeatother

\usepackage{hyperref}
\usepackage[sort&compress,numbers,merge]{natbib}
\usepackage{mmacells}
\mmaSet{
  morefv={gobble=2},
  linklocaluri=mma/symbol/definition:#1,
  morecellgraphics={yoffset=1.9ex},
  moredefined={
  NiceForm, FreeLag, e, A, y, Bar, PR,
  Match, EFTorder, LoopOrder,
  HcSimplify, CollectTerms, EoMSimplify, IBPSimplify
  }
}

\title{Towards matching effective theories efficiently\,\footnote{Contribution to the ``Les Rencontres de Physique de la Vall\'ee d'Aoste'', La~Thuile, 2022.}}
\author{
F.~Wilsch
\\[0.5cm]
\it 
Physik-Institut, Universit\"at Z\"urich, CH-8057 Z\"urich, Switzerland
\\[0.2cm]
}

\begin{document}
\renewcommand*{\thefootnote}{\fnsymbol{footnote}}
\maketitle
\setcounter{footnote}{0}
\renewcommand*{\thefootnote}{\arabic{footnote}}

% ABSTRACT
\begin{abstract}
\noindent
Separation of scales in quantum field theories is essential when studying the low-energy phenomenology of a given UV~model. To this end, it is necessary to construct an effective field theory containing only the light degrees of freedom and matching it to the full theory, ensuring that both describe the same low-energy dynamics. Performing this matching beyond the leading order is crucial, as a great number of observables, like FCNC, only appear at the loop level in the Standard Model and in many new physics scenarios. One possibility to obtain the low-energy theory is by integrating out the heavy particles from the full theory using path integral techniques. We review this functional matching procedure at the one-loop level and discuss common challenges involved in determining the matching conditions for effective theories. Due to the great diversity of beyond the Standard Model theories and the complexity of the matching computations, an automation of this procedure is desirable. On this matter, we present the ongoing effort to develop the \texttt{Mathematica} package {\tt Matchete} facilitating the fully automatic matching for a broad range of theories. When completed, this will significantly simplify the analysis of the low-energy phenomenology of beyond the Standard Model physics. 
These proceedings are based on refs.~\cite{Fuentes-Martin:2020udw} and~\cite{Matchete}.
\end{abstract}

\section{Introduction}
Effective Field Theory~(EFT) is a powerful tool to study the low-energy dynamics of quantum field theories involving particles of different mass scales. Consider a theory containing heavy particles~$\eta_H$ with masses $m_H$ and light particles~$\eta_L$ with masses~$m_L$ with a large scale separation~$m_L \ll m_H$. If an experiment is performed at low energies~$\smash{E \sim \sqrt{q^2} \sim m_L}$, where $q$ is the typical momentum scale of the light particles in that process, the heavy degrees of freedom do not contribute as on-shell states, and thus their effects can be considered as sub-leading corrections to the measurements. In this scenario, due to the resummation of large logarithms, it is required to construct an EFT containing only the light particles, where the effects of the heavy particles are encoded in a tower of higher-dimensional operators built out of $\eta_L$ fields, that are suppressed by appropriate powers of~$1/m_H$. An example for such a model is Fermi's theory, where the heavy particle removed from the theory is the $W$~boson, and the interactions at low energies are described by four-fermion contact interactions that are suppressed by~$1/m_W^{2}$.

There are two main challenges for the construction of EFTs: $i)$~Finding all higher-dimensional operators that have to be considered such that the EFT can describe the same low-energy phenomenology as the corresponding UV theory; $ii)$~Determining the values of the coefficients of all effective operators in terms of the UV Lagrangian parameters. The latter is often referred to as \textit{matching} the EFT to the~UV. Traditionally, this is done by computing all Feynman diagrams that contribute to the effective action~$\Gamma[\eta_L]$, as a function of the light fields only, in both theories and equating these $\Gamma_\mathrm{EFT}=\Gamma_\mathrm{UV}$, where $\Gamma_\mathrm{UV}$ has to be expanded as a power series in~$1/m_H$ (see \textit{e.g.} \cite{Georgi:1991ch}). This method is called \textit{diagrammatic matching}. In the following, however, we will consider a different approach to EFT matching based on the path integral. When we construct an EFT it is often important to do this at the loop level, meaning we have to consider effective operators whose coefficients get a non-vanishing value when the matching is performed at the loop level. The reason is that many interesting phenomena, such as FCNCs, do not appear at tree level in the Standard Model~(SM) and many New Physics~(NP) scenarios.

Effective theories also play a crucial role in the analysis of Beyond-the-SM~(BSM) theories. The absence of direct signals of NP at the experiments at the LHC indicate that the mass scale of the potential new particles is far above the electroweak scale~$v \ll m_H$, although light NP is not fully excluded. In the case of heavy NP, the effects of the BSM particles can be described at lower energies by the so-called Standard Model Effective Field Theory~(SMEFT)~\cite{Buchmuller:1985jz,Grzadkowski:2010es}, which allows to study the NP effects at the electroweak scale~$v$ in a model independent way. 
To analyze the effects of a UV theory at even lower energies, we can first match it to the SMEFT, then use the renormalization group equations~(RGE) of the SMEFT~\cite{SMEFT_RGE} to evolve the coefficients to lower scales, and then match the SMEFT to another EFT, often called LEFT~\cite{Jenkins:2017jig}, where the heaviest particles of the SM (\textit{i.e.} the top quark, $W/Z$~boson, and the Higgs) are removed. This EFT is then appropriate to compare the theory to the low-energy experiments.

These multi-step matching scenarios can be computationally very challenging, as all matching steps have to be performed at the loop level, and the RGE of all EFTs need to be determined. Therefore, an automation of this procedure is important to allow for analyses of the vast variety of different BSM theories. In these proceeding, which are based on refs.~\cite{Fuentes-Martin:2020udw} and~\cite{Matchete}, we review the functional method for EFT matching developed, \textit{e.g.}, in refs.~\cite{Henning:2014wua,delAguila:2016zcb,Henning:2016lyp,Fuentes-Martin:2016uol,Zhang:2016pja,Cohen:2020fcu}, which both derives the effective operators and their matching conditions. The method is particularly suited for an automation and we present the ongoing effort in the development of the \texttt{Mathematica} code \texttt{Matchete}~\cite{Matchete} -- matching effective theories efficiently -- which will, when released, fully automate the EFT matching at one loop, thus significantly simplifying the analysis of BSM theories.

% % % % % % % % % % % % % % % % % % % % % % % % % % % % % % % % % % % % % % % % % % % % % % % % % % % % % % %
\section{Functional methods for EFT matching}
\label{sec:functional-method}
In this section we review the functional matching procedure based on the path integral formalism and the background field method, for which we will mostly follow refs.~\cite{Fuentes-Martin:2020udw,Cohen:2020fcu}.

%was developed in refs.~\cite{Henning:2014wua,delAguila:2016zcb,Henning:2016lyp,Fuentes-Martin:2016uol,Zhang:2016pja,Cohen:2020fcu}

\subsection{Functional formalism and tree-level matching}
Consider a general UV theory described by the Lagrangian~$\mathcal{L}_\mathrm{UV}$ containing both heavy $\eta_H$ and light $\eta_L$ particles with a mass gap~$m_H \gg m_L$. The corresponding fields can be grouped in a multiplet ${\eta=(\eta_H,\,\eta_L)}$. Using the background field method we split all fields by $\eta\to\hat\eta+\eta$, where $\hat\eta$ is the background field configuration satisfying the classical equations of motion~(EOM), and $\eta$ are the quantum fluctuations. In Feynman diagrams tree-level lines thus arise from the background fields~$\hat\eta$, whereas lines in loops arise from the quantum fluctuations~$\eta$. Therefore, to capture all effects up to one loop, it is sufficient to expand the Lagrangian up to quadratic order in the fluctuations~$\eta$
\begin{eqnarray}
\label{eq:Lagrangian}
\mathcal{L}_\mathrm{UV}[\hat\eta+\eta]
&= \mathcal{L}_\mathrm{UV}[\hat\eta] + \frac{1}{2} \bar{\eta}_i \left. \frac{\delta^2 \mathcal{L}_\mathrm{UV}}{\delta \bar\eta_i \, \delta\eta_j} \right|_{\eta=\hat\eta} \eta_j + \mathcal{O}(\eta^3) \,, 
\end{eqnarray}
where the functional derivative with respect to the quantum fluctuations is denoted~$\delta/\delta\eta_i$. The linear terms in the expansion above vanish due to the EOM, and higher-order terms only contribute at higher loop orders.

The effective action of the theory can be expressed as a path integral by
\begin{eqnarray}
\label{eq:effective_action}
\exp\left( i \Gamma_\mathrm{UV}[\hat\eta] \right)
&= \int \mathcal{D}\eta \, \exp\left( i\int\mathrm{d}^dx\,\mathcal{L}_\mathrm{UV}[\hat\eta+\eta] \right)\,.
\end{eqnarray}
Combining eqs.~\eqref{eq:Lagrangian} and~\eqref{eq:effective_action} we find the tree-level effective action $\smash{\Gamma_\mathrm{UV}^{(0)}[\hat\eta] = \int \mathrm{d}^d x \, \mathcal{L}_\mathrm{UV}[\hat\eta]}$.
At energies far below the heavy mass, \textit{i.e.} for $q^2\sim m_L^2 \ll m_H^2$, we can solve the EOM for $\hat{\eta}_H$ by expanding in inverse powers of~$m_H$, thus obtaining a solution of the form $\hat\eta_H=\hat\eta_H(\hat\eta_L)$ as a perturbative series in $1/m_H$. This yields the tree-level EFT Lagrangian
\begin{eqnarray}
\mathcal{L}_\mathrm{EFT}^{(0)}[\hat\eta_L] &= \mathcal{L}_\mathrm{UV}[\hat\eta_L,\hat\eta_H(\hat\eta_L)] \,,
\end{eqnarray}
which only depends on light background fields.

\subsection{One-loop matching}
Similarly the one-loop effective action is given by
\begin{eqnarray}
\exp\left( i \Gamma_\mathrm{UV}^{(1)} \right)
&= \int\mathcal{D}\eta \, \exp\left( \frac{i}{2} \int\mathrm{d}^d x \, \bar{\eta}_i \mathcal{O}_{ij} \eta_j \right) \,, \quad \mathrm{where} \quad \mathcal{O}_{ij} = \left .\frac{\delta^2 \mathcal{L}_\mathrm{UV}}{\delta \bar{\eta}_i \, \delta{\eta}_j} \right|_{\eta=\hat\eta} \,.
\end{eqnarray}
The operator $\mathcal{O}_{ij}$ is dubbed the fluctuation operator. The one-loop effective action is thus determined by a Gaussian path integral which can be readily solved
\begin{eqnarray}
\Gamma_\mathrm{UV}^{(1)} 
&= \frac{i}{2} \log \mathrm{SDet} \, \mathcal{O}
= \frac{i}{2} \mathrm{STr} \log \mathcal{O} \,.
\end{eqnarray}
Here, we introduced the superdeterminant~SDet and the supertrace~STr, which are generalizations of the determinant and the trace, respectively, to matrices with both Grassmann and ordinary elements, where, \textit{e.g.}, STr carries opposite sign for fermionic and bosonic entries.
The supertrace is a trace over all internal degrees of freedom including momenta, and thus also contains a loop integral
\begin{eqnarray}
\label{eq:loop-integral}
\mathrm{STr} \log \mathcal{O}
&=
\pm \int \frac{\mathrm{d}^d k}{(2\pi)^d} \langle k | \mathrm{tr}\log\mathcal{O} | k \rangle \,,
\end{eqnarray}
where to regulate divergences we use dimensional regularization with the $\overline{\mathrm{MS}}$-scheme.

We split the fluctuation operator into a propagator~$\Delta_i$ and an interaction part~$X_{ij}$
\begin{align}
\label{eq:fluctuation-operator}
\mathcal{O}_{ij} &= \delta_{ij} \Delta_i^{-1} - X_{ij}
= \Delta_{ij}^{-1} \left( \delta_{ij} - \Delta_i X_{ij} \right) \, , 
& 
&\mathrm{where}
&
\Delta_{i}^{-1} &= \left\{\begin{matrix}
	-(D^2+m_i^2)
	\\
	i \slashed{D} - m_i
	\\
	g^{\mu\nu}(D^2+m_i^2)
\end{matrix}
\right.
\end{align}
for scalars, fermions and vector bosons, respectively.
For simplicity, we work in the Feynman gauge for the propagators of the quantum fluctuations. Notice that this does not imply a particular gauge choice for the light background fields which remain in a general $R_\xi$~gauge (see ref.~\cite{Henning:2014wua} for more details). 
%It is convenient to treat light masses~$m_L$ as interactions and include them in the $X_{ij}$~terms, thus setting $m_i=0$ for light fields in the propagators~$\Delta_{i}$. 
The one-loop effective action thus reads
\begin{eqnarray}
\Gamma_\mathrm{UV}^{(1)} 
&= \frac{i}{2} \mathrm{STr} \log \Delta^{-1} + \frac{i}{2} \mathrm{STr} \log \left(1-\Delta X\right) \,.
\label{eq:on-loop_effective_action}
\end{eqnarray}

To further simplify~$\Gamma_\mathrm{UV}^{(1)}$ we apply the method of \textit{expansion by regions}~\cite{Beneke:1997zp,Jantzen:2011nz}. The loop integrals can depend in principle on the scales~$m_H$, $m_L$, and~$q$, where all external momenta~$q$ are assumed to be of order~$q^2\sim m_L^2$. Thus we separate all loop integrals coming from eq.~\eqref{eq:loop-integral} into a hard and soft region, where the loop momentum~$k$ is assumed to be hard~$k\sim m_H$ and soft~$k\sim m_L$, respectively. We then expand each region in the quantities that are small, \textit{i.e.} $m_L/k$ for the hard region and $k/m_H$ in the soft region, and consequently integrate both regions over the full $d$-dimensional space of loop momenta. 
Summing both expanded integrals then yields the full original loop integral.

Following this procedure we can write
\begin{eqnarray}
\Gamma_\mathrm{UV}^{(1)} &= \left. \Gamma_\mathrm{UV}^{(1)} \right|_\mathrm{hard} + \left. \Gamma_\mathrm{UV}^{(1)} \right|_\mathrm{soft}.
\end{eqnarray}
An essential simplification arrises by realizing that $\smash{\Gamma_\mathrm{UV}^{(1)} \big|_\mathrm{soft}}$ contains the same contributions as one-loop contributions by the tree-level effective Lagrangian~$\smash{\mathcal{L}_\mathrm{EFT}^{(0)}}$. Therefore, the one-loop contributions to the EFT Wilson coefficients in~$\smash{\mathcal{L}_\mathrm{EFT}^{(1)}}$ are entirely encoded in the hard region of the effective action
\begin{eqnarray}
\left. \Gamma_\mathrm{UV}^{(1)} \right|_\mathrm{hard}
&=
\int \mathrm{d}^d x \, \mathcal{L}_\mathrm{EFT}^{(1)} \,.
\label{eq:one-loop_lagrangian}
\end{eqnarray}
Thus, only supertraces with at least one heavy propagator in the loop can contribute to~$\mathcal{L}_\mathrm{EFT}^{(1)}$, since loops without heavy propagators only contribute to the soft region.

Combining our results in eqs.~\eqref{eq:on-loop_effective_action} and~\eqref{eq:one-loop_lagrangian} we obtain the one-loop EFT matching condition% for the functional procedure
\begin{eqnarray}
\int\mathrm{d}^d x \, \mathcal{L}_\mathrm{EFT}^{(1)}
&=
\frac{i}{2} \left. \mathrm{STr} \log \Delta^{-1} \right|_\mathrm{hard}
- \frac{i}{2} \sum_{n=0}^{\infty} \frac{1}{n} \left. \mathrm{STr}\left[\left(\Delta X\right)^n\right] \right|_\mathrm{hard} \,,
\label{eq:master}
\end{eqnarray}
where we expanded the logarithm in the last term since in the hard region $\Delta X$ is at most~$\mathcal{O}(m_H^{-1})$. We find that $\smash{\mathcal{L}_\mathrm{EFT}^{(1)}}$ is obtained by computing two types of supertraces:
\begin{itemize}
\item \textit{Log-type supertraces} only depend on heavy particle propagators~$\Delta_i$ as the light particles only contribute to the soft region. Furthermore, they are model independent as they do not depend on the interactions~$X_{ij}$. 
\item \textit{Power-type supertraces} contain contributions by both heavy and light fields, but at least one of the propagators in $(\Delta X)^n$ must correspond to a heavy particle for it to contribute to the hard region. Since $\Delta X$ is at most $\mathcal{O}(m_H^{-1})$ only a finite number of STr has to be computed in eq.~\eqref{eq:master} when working up to a fixed EFT~order.
\end{itemize}

The supertraces can be evaluated in a manifestly gauge covariant form using a \textit{covariant derivative expansion}~\cite{Henning:2014wua,Henning:2016lyp}. A detailed description of the calculation of supertraces is beyond the scope of these proceedings and we refer to refs.~\cite{Fuentes-Martin:2020udw,Cohen:2020fcu,Cohen:2020qvb} for more details.

%For more details on the evaluation of the supertraces see refs.~\cite{Fuentes-Martin:2020udw} \textsc{More To Be Added!}

% % % % % % % % % % % % % % % % % % % % % % % % % % % % % % % % % % % % % % % % % % % % % % % % % % % % % % %
\section{\texttt{Matchete} -- matching effective theories efficiently}
Due to the great variety of UV models, and due to the complexity of the supertrace evaluation for a particular model, an automation is required. In recent years there was an increased effort in the community to built tools automating at least partially the EFT matching~\cite{Fuentes-Martin:2020udw, Carmona:2021xtq, Criado:2017khh, Cohen:2020qvb, DasBakshi:2018vni}. 
Here we show a preview of the \texttt{Mathematica} package \texttt{Matchete}~\cite{Matchete}, that we are currently developing. \texttt{Matchete} implements the functional matching procedure outlined in sec.~\ref{sec:functional-method} and is built upon \texttt{SuperTracer}~\cite{Fuentes-Martin:2020udw} which handles the identification and evaluation of the suptertraces. However, when made public, \texttt{Matchete} will offer further functionalities such as automatic tree-level matching, derivation of the fluctuation operator, and EFT operator simplification, providing a complete software suite for fully automatic one-loop matching.

The functional matching procedure is well suited for this task, as it can be applied to generic weakly coupled UV theories, which have a mass power counting that allows for the construction of the corresponding low-energy effective theory. Thus, \texttt{Matchete} can be used with a wide range of theories that can, but do not necessarily have to, match to the SMEFT. Furthermore, the matching can be performed to in principle arbitrary EFT dimension. However, in practice the procedure is limited by computation time. 

\begin{figure}[]
\centering
\includegraphics[width=.95\textwidth]{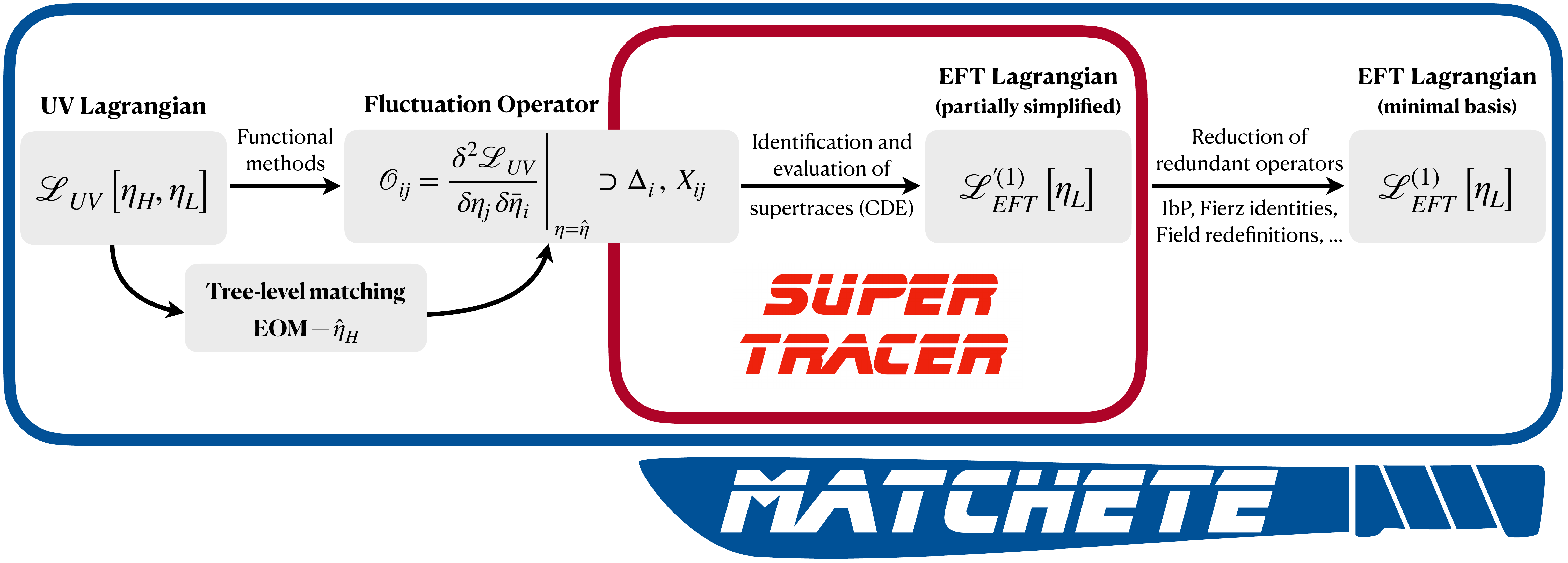}    % includes figure foo.eps
\caption{Illustration of the functional EFT matching procedure. The blue (red) box show the parts of the procedure that are handled by the \texttt{Matchete}~\cite{Matchete} (\texttt{SuperTracer}~\cite{Fuentes-Martin:2020udw}) program.}
\label{fig:workflow}
\end{figure}

The workflow of \texttt{Matchete} is illustrated in fig.~\ref{fig:workflow} and a working example is presented in sec.~\ref{sec:example}. The starting point is a UV Lagrangian~$\mathcal{L}_\mathrm{UV}$ that has to be provided by the user. This must include the specification of all symmetries, the field content, the interactions, and the power counting of the theory. Starting from this input and by applying functional derivatives on~$\mathcal{L}_\mathrm{UV}$, the code derives and solves the EOMs for the heavy fields, thus obtaining the tree-level EFT Lagrangian. The program then derives the fluctuation operator~$\mathcal{O}_{ij}$, and with that all propagators~$\Delta_i$ and interaction terms~$X_{ij}$ of the theory, as defined in eq.~\eqref{eq:fluctuation-operator}. These are then passed on to the routines of the \texttt{SuperTracer}~\cite{Fuentes-Martin:2020udw} package,\,\footnote{\texttt{SuperTracer} will be fully integrated into \texttt{Matchete}, and not be available as a standalone after the release of \texttt{Matchete}.} which identify and evaluate all relevant supertraces. 

The one-loop EFT Lagrangian thus obtained includes, in general, redundant operators, \textit{i.e.} operators that can be related to other operators present in the Lagrangian by operations like: integration by parts, field redefinitions, Fierz identities, Dirac algebra identities, etc. These identities and relations are implemented in \texttt{Matchete}, allowing for a further simplification of the Lagrangian obtained from the supertraces. The eventual goal is to obtain the EFT Lagrangian in a minimal basis without any redundancies.

To our knowledge, so far there is only one other fully automatic matching tool available called \texttt{Matchmakereft}~\cite{Carmona:2021xtq}. This code performs the matching of effective theories based on the conventional diagrammatic matching procedure.
Since \texttt{Matchmakereft} and \texttt{Matchete} are based on different techniques, having two programs available will allow for a more precise validation of both, although \texttt{Matchmakereft} and \texttt{SuperTracer} have already been checked against results available in the literature such as, \textit{e.g.}, refs.~\cite{Gherardi:2020det,Zhang:2021jdf}.

A well-known advantage of the functional matching prescription is that the EFT operators are directly obtained by evaluating the supertraces in eq.~\eqref{eq:master}. Thus, contrary to the diagrammatic matching procedure, no a~priori knowledge of an EFT basis is required for the matching calculation. This can significantly simplify the matching in the case where the UV theory does not match onto an EFT with a well known basis for matching, such as the SMEFT, since the construction of a basis can be a non-trivial~task. 
%Furthermore, the functional matching procedure offers a manifestly covariant prescription for EFT matching through the \textit{covariant derivative expansion}~\cite{Henning:2014wua,Henning:2016lyp}.

% % % % % % % % % % % % % % % % % % % % % % % % % % % % % % % % % % % % % % % % % % % % % % % % % % % % % % %
\section{One-loop matching of a vectorlike fermion toy model}
\label{sec:example}
In this section we reconsider a toy model with vectorlike fermions, already discussed in ref.~\cite{Fuentes-Martin:2020udw}, to illustrate the functional matching procedure. However, here we will focus on the application of \texttt{Matchete} to this example. Notice that the code examples shown below are preliminary and can be subject to change in the version published eventually.

Consider a theory with a $U(1)$~gauge symmetry, where the gauge bosons is labeled~$A_\mu$. In addition we include two vectorlike fermions that both have charge~$1$ under the $U(1)$ group. One of the fermions is heavy~$\Psi$ with mass~$M$, whereas the other fermion~$\psi$ is massless. Furthermore, we add a massless real scalar singlet~$\phi$ to the theory that interacts with the fermions through a Yukawa interaction with coupling~$y$. The Lagrangian of this UV theory reads
\begin{align}
\label{eq:example_Lagrangian}
\mathcal{L}_\mathrm{UV}
&= 
-\frac{1}{4} A_{\mu\nu} A^{\mu\nu} 
+ \frac{1}{2} (\partial_\mu \phi)(\partial^\mu \phi)
+ i \bar{\psi} \slashed{D} \psi
+ i \bar{\Psi} \slashed{D} \Psi - M \bar{\Psi}\Psi
- (y \bar{\psi}_L \phi \Psi_R + \mathrm{h.c.})
\,,
\end{align}
with the covariant derivative $D_\mu = \partial_\mu - i g A_\mu$, and the field-strength tensor $A_{\mu\nu}=\partial_\mu A_\nu - \partial_\nu A_\mu$. This theory is specified in \texttt{Matchete} with the commands outlined below.

We can define the $U(1)$~gauge group labeled \mmaInlineCell[]{Input}{U1e} by
\begin{mmaCell}{Input}
  \mmaDef{DefineGaugeGroup}[U1e, \mmaDef{U}[1], e, A]
\end{mmaCell}
which also defines the associated coupling~\mmaInlineCell[]{Input}{e}, gauge field and field-strength tensor labeled~\mmaInlineCell[]{Input}{A}. The remaining fields can then be defined with the commands
\begin{mmaCell}{Input}
  \mmaDef{DefineField}[\(\Psi\), \mmaDef{Fermion}, \mmaDef{Charges} -> \{U1e[1]\}, \mmaDef{Mass} -> \{\mmaDef{Heavy}, \mmaDef{M}\}]
  \mmaDef{DefineField}[\(\psi\), \mmaDef{Fermion}, \mmaDef{Charges} -> \{U1e[1]\}, \mmaDef{Mass} -> 0]
  \mmaDef{DefineField}[\(\phi\), \mmaDef{Scalar}, \mmaDef{Mass} -> 0, \mmaDef{SelfConjugate} -> True]
\end{mmaCell}
Finally, we define the Yukawa coupling \mmaInlineCell[]{Input}{y}, which is of order~$\mathcal{O}(m_L^0)$, by
\begin{mmaCell}{Input}
  \mmaDef{DefineCoupling}[y, \mmaDef{EFTorder} -> 0]
\end{mmaCell}
After these specifications \texttt{Matchete} can automatically construct the free Lagrangian of the theory with the routine \mmaInlineCell[]{Input}{FreeLag} which needs to be combined with the Yukawa interactions to obtain the full Lagrangian shown in eq.~\eqref{eq:example_Lagrangian}
\begin{mmaCell}{Input}
  \mmaDef{LUV} = FreeLag[] - \mmaDef{PlusHc}[y[] \(\phi\)[] Bar[\(\psi\)[]]**PR**\(\Psi\)[]];
  \mmaDef{LUV} // NiceForm
\end{mmaCell}
\begin{mmaCell}{Output}
  -\(\frac{1}{4}\)\mmaSup{\mmaSup{A}{\(\mu\nu\)}}{2}+\(\frac{1}{2}\)\mmaSup{(\mmaSub{D}{\(\mu\)}\(\phi\))}{2}+i(\(\bar{\psi}\)\(\cdot\)\mmaSub{\(\gamma\)}{\(\mu\)}\(\!\cdot\)\mmaSub{D}{\(\mu\)}\(\cdot\)\(\psi\))+i(\(\bar{\Psi}\)\(\cdot\)\mmaSub{\(\gamma\)}{\(\mu\)}\(\!\cdot\)\mmaSub{D}{\(\mu\)}\(\cdot\)\(\Psi\))-M(\(\bar{\Psi}\)\(\cdot\)\(\Psi\))-\(\big(\)y\(\phi\)(\(\bar{\psi}\)\(\cdot\)\mmaSub{P}{R}\(\cdot\)\(\Psi\))+h.c.\(\!\big)\)
\end{mmaCell}
%+\mmaOver{y}{_}\(\phi\)(\(\bar{\Psi}\)\(\cdot\)\mmaSub{P}{L}\(\cdot\)\(\psi\))
where \mmaInlineCell[]{Input}{\mmaDef{PR}} is a right-chirality projector, the Hermitian conjugate is included through \mmaInlineCell[]{Input}{\mmaDef{PlusHc}}, and all objects carrying spinor indices must be contracted using \mmaInlineCell[]{Input}{\mmaDef{**}}, due to their non-commutativity. In the second line the Lagrangian is printed in a readable format using the \mmaInlineCell[]{Input}{NiceForm} routine.

The dimension-six tree-level EFT Lagrangian can be derived with the \mmaInlineCell[]{Input}{Match} routine
\begin{mmaCell}{Input}
  \mmaDef{LEFT0} = Match[\mmaDef{LUV}, LoopOrder -> 0, EFTorder -> 6];
  \mmaDef{LEFT0} // NiceForm
\end{mmaCell}
\begin{mmaCell}{Output}
  -\(\frac{1}{4}\)\mmaSup{\mmaSup{A}{\(\mu\nu\)}}{2}+\(\frac{1}{2}\)\mmaSup{(\mmaSub{D}{\(\mu\)}\(\phi\))}{2}+i(\(\bar{\psi}\)\(\cdot\)\mmaSub{\(\gamma\)}{\(\mu\)}\(\cdot\)\mmaSub{D}{\(\mu\)}\(\cdot\)\(\psi\))+i\mmaFrac{\mmaOver{y}{_}y}{\mmaSup{M}{2}}\(\phi\)\mmaSub{D}{\(\mu\)}\(\phi\)(\(\bar{\psi}\)\(\cdot\)\mmaSub{\(\gamma\)}{\(\mu\)}\mmaSub{P}{L}\(\cdot\)\(\psi\))+i\mmaFrac{\mmaOver{y}{_}y}{\mmaSup{M}{2}}\mmaSup{\(\phi\)}{2}(\(\bar{\psi}\)\(\cdot\)\mmaSub{\(\gamma\)}{\(\mu\)}\mmaSub{P}{L}\(\cdot\)\mmaSub{D}{\(\mu\)}\(\cdot\)\(\psi\))
\end{mmaCell}
The contribution by the last two terms vanishes on-shell, as can be shown using integration by parts and field redefinitions. These redundancies are removed in \texttt{Matchete}~by
\begin{mmaCell}{Input}
  \mmaDef{LEFT0min} = EoMSimplify[IBPSimplify[\mmaDef{LEFT0}]];
  \mmaDef{LEFT0min} // NiceForm
\end{mmaCell}
\begin{mmaCell}{Output}
  -\(\frac{1}{4}\)\mmaSup{\mmaSup{A}{\(\mu\nu\)}}{2} +\(\frac{1}{2}\)\mmaSup{(\mmaSub{D}{\(\mu\)}\(\phi\))}{2} +i(\(\bar{\psi}\)\(\cdot\)\mmaSub{\(\gamma\)}{\(\mu\)}\(\cdot\)\mmaSub{D}{\(\mu\)}\(\cdot\)\(\psi\))
\end{mmaCell}
where the EFT contribution vanishes since $\psi$ is massless leading to the equation of motion~$\slashed{D}\psi=0$.
Integrating out the heavy fermion~$\Psi$ at one loop and dimension six in the EFT, can be achieved similarly
\begin{mmaCell}{Input}
  \mmaDef{LEFT1} = Match[\mmaDef{LUV}, LoopOrder -> 1, EFTorder -> 6];
  \mmaDef{LEFT1min} = EoMSimplify[IBPSimplify[\mmaDef{LEFT1}]];
  \mmaDef{LEFT1min} // NiceForm
\end{mmaCell}
\begin{mmaCell}{Output}
  -\mmaFrac{1}{4}\mmaSup{\mmaSup{A}{\(\mu\nu\)}}{2} +\mmaFrac{1}{2}\mmaSup{(\mmaSub{D}{\(\mu\)}\(\phi\))}{2} +i(\mmaOver{\(\psi\)}{_}\(\cdot\)\mmaSub{\(\gamma\)}{\(\mu\)}\(\cdot\)\mmaSub{D}{\(\mu\)}\(\psi\))
  -2\(\hbar\)\mmaOver{y}{_}y\mmaSup{M}{2}\Big(1+Log\Big[\mmaFrac{\mmaSup{\(\mu\)}{2}}{\mmaSup{M}{2}}\Big]\Big)\mmaSup{\(\phi\)}{2} -\(\hbar\)\mmaSup{\mmaOver{y}{_}}{2}\mmaSup{y}{2}Log\Big[\mmaFrac{\mmaSup{\(\mu\)}{2}}{\mmaSup{M}{2}}\Big]\mmaSup{\(\phi\)}{4} +\mmaFrac{1}{3}\(\hbar\)\mmaFrac{\mmaSup{\mmaOver{y}{_}}{3}\mmaSup{y}{3}}{\mmaSup{M}{2}}\mmaSup{\(\phi\)}{6} +\mmaFrac{1}{3}\(\hbar\)\mmaFrac{\mmaOver{y}{_}y\mmaSup{e}{2}}{\mmaSup{M}{2}}\mmaSup{\(\phi\)}{2}\mmaSup{\mmaSup{A}{\(\mu\nu\)}}{2}
  -\mmaFrac{2}{15}\(\hbar\)\mmaFrac{\mmaSup{e}{4}}{\mmaSup{M}{2}}\mmaSup{(\mmaOver{\(\psi\)}{_}\(\cdot\)\mmaSub{\(\gamma\)}{\(\mu\)}\(\cdot\psi\))}{2} +\mmaFrac{7}{36}\(\hbar\)\mmaFrac{\mmaOver{y}{_}y\mmaSup{e}{2}}{\mmaSup{M}{2}}(\mmaOver{\(\psi\)}{_}\(\cdot\)\mmaSub{\(\gamma\)}{\(\mu\)}\(\cdot\psi\))(\mmaOver{\(\psi\)}{_}\(\cdot\)\mmaSub{\(\gamma\)}{\(\mu\)}\mmaSub{P}{L}\(\cdot\psi\))
\end{mmaCell}
where we directly applied the simplification routines. Here $\hbar$ represents a loop factor, \textit{i.e.} $\hbar\to\frac{\hbar}{16\pi^2}$. This gives the final matching result for the EFT Lagrangian at one loop. The results of this computation have been partially cross checked in ref.~\cite{Fuentes-Martin:2020udw} with a diagrammatic by hand calculation. The automatic implementation, however, is considerably less time consuming and also less prone to errors than the derivation by hand, showing clearly the advantages of using codes such as \texttt{Matchete} for one-loop matching computations, even for fairly simple models such as the one considered here.

% % % % % % % % % % % % % % % % % % % % % % % % % % % % % % % % % % % % % % % % % % % % % % % % % % % % % % %
\section{Conclusions}
The construction of EFTs describing the same low-energy dynamics as a given NP model is a common problem when studying BSM theories. Since EFT matching often has to be performed at the one-loop level and even several matching steps might be involved, this derivation is a challenging although very technical task. Thus, an automated solution for these calculations would significantly simplify the analyses of possible BSM theories. In these proceeding we presented and summarized the functional matching procedure, in which the heavy particles of a theory are removed from the theory by integrating over them in the path integral. This method is particularly suited for an automation in a computer algebra code, as the path integral yields both the effective operators and their coefficients, removing the need to first derive the EFT operator basis before starting the matching calculation, as is necessary for the diagrammatic matching procedure. 
The functional method is currently being implemented in the \texttt{Mathematica} code \texttt{Matchete} and we illustrated the functionalities of this package, by performing the fully automatic one-loop matching for a toy model with vectorlike fermions in this talk. When released, \texttt{Matchete} will significantly simplify the analysis of a wide range of different BSM~theories.

% % % % % % % % % % % % % % % % % % % % % % % % % % % % % % % % % % % % % % % % % % % % % % % % % % % % % % %
\section*{Acknowledgments}
I would like to thank the organizers of the ``Les Rencontres de Physique de la Vall\'ee d'Aoste'' for the invitation. I am very grateful to my colleagues Javier Fuentes-Mart\'in, Matthias K\"onig, Julie Pag\`es, and Anders E. Thomsen for their collaboration on this project.
The author has received funding from the European Research Council (ERC) under the European Union's Horizon 2020 research and innovation program under grant agreement 833280 (FLAY). The participation in the conference was supported by a scholarship awarded by the~INFN.

% - - - - Bibliography - - - - %


\footnotesize
\begin{thebibliography}{0}

%\cite{Fuentes-Martin:2020udw}
\bibitem{Fuentes-Martin:2020udw}
J.~Fuentes-Martin, M.~K\"onig, J.~Pag\`es, A.~E.~Thomsen and F.~Wilsch,
%``SuperTracer: A Calculator of Functional Supertraces for One-Loop EFT Matching,''
JHEP \textbf{04}, 281 (2021).
%doi:10.1007/JHEP04(2021)281
%[arXiv:2012.08506].

\bibitem{Matchete}
J.~Fuentes-Martin, M.~K\"onig, J.~Pag\`es, A.~E.~Thomsen and F.~Wilsch,
%``SuperTracer: A Calculator of Functional Supertraces for One-Loop EFT Matching,''
\textit{in preparation}.

%\cite{Georgi:1991ch}
\bibitem{Georgi:1991ch}
H.~Georgi,
%``On-shell effective field theory,''
Nucl. Phys. B \textbf{361}, 339-350 (1991),
%doi:10.1016/0550-3213(91)90244-R
%256 citations counted in INSPIRE as of 04 Jun 2022
%\cite{Georgi:1993mps}
%\bibitem{Georgi:1993mps}
%H.~Georgi,
%``Effective field theory,''
Ann.\,Rev.\,Nucl.\,Part.\,Sci.\,\textbf{43},\,209-252\,(1993).
%doi:10.1146/annurev.ns.43.120193.001233
%577 citations counted in INSPIRE as of 04 Jun 2022

%\cite{Buchmuller:1985jz}
\bibitem{Buchmuller:1985jz}
W.~Buchmuller and D.~Wyler,
%``Effective Lagrangian Analysis of New Interactions and Flavor Conservation,''
Nucl. Phys. B \textbf{268}, 621-653 (1986).
%doi:10.1016/0550-3213(86)90262-2
%2056 citations counted in INSPIRE as of 02 Jun 2022

%\cite{Grzadkowski:2010es}
\bibitem{Grzadkowski:2010es}
B.~Grzadkowski, M.~Iskrzynski, M.~Misiak and J.~Rosiek,
%``Dimension-Six Terms in the Standard Model Lagrangian,''
JHEP \textbf{10}, 085 (2010).
%doi:10.1007/JHEP10(2010)085
%[arXiv:1008.4884].
%1667 citations counted in INSPIRE as of 02 Jun 2022

%\cite{SMEFT_RGE}
\bibitem{SMEFT_RGE}
R.~Alonso, E.~E.~Jenkins, A.~V.~Manohar and M.~Trott,
JHEP \textbf{10}, 087 (2013),
%[arXiv:1308.2627]
JHEP \textbf{01}, 035 (2014),
%[arXiv:1310.4838]
JHEP \textbf{04}, 159 (2014).
%[arXiv:1312.2014].

%\cite{Jenkins:2017jig}
\bibitem{Jenkins:2017jig}
E.~E.~Jenkins, A.~V.~Manohar and P.~Stoffer,
%``Low-Energy Effective Field Theory below the Electroweak Scale: Operators and Matching,''
JHEP \textbf{03}, 016 (2018).
%doi:10.1007/JHEP03(2018)016
%[arXiv:1709.04486].
%141 citations counted in INSPIRE as of 02 Jun 2022

%\cite{Cohen:2020fcu}
\bibitem{Cohen:2020fcu}
T.~Cohen, X.~Lu and Z.~Zhang,
%``Functional Prescription for EFT Matching,''
JHEP \textbf{02}, 228 (2021).
%doi:10.1007/JHEP02(2021)228
%[arXiv:2011.02484].
%17 citations counted in INSPIRE as of 01 Jun 2022

%\cite{Henning:2014wua}
\bibitem{Henning:2014wua}
B.~Henning, X.~Lu and H.~Murayama,
%``How to use the Standard Model effective field theory,''
JHEP \textbf{01}, 023 (2016).
%doi:10.1007/JHEP01(2016)023
%[arXiv:1412.1837].
%249 citations counted in INSPIRE as of 01 Jun 2022

%\cite{delAguila:2016zcb}
\bibitem{delAguila:2016zcb}
F.~del Aguila, Z.~Kunszt and J.~Santiago,
%``One-loop effective lagrangians after matching,''
Eur. Phys. J. C \textbf{76}, no.5, 244 (2016).
%doi:10.1140/epjc/s10052-016-4081-1
%[arXiv:1602.00126].
%80 citations counted in INSPIRE as of 01 Jun 2022

%\cite{Henning:2016lyp}
\bibitem{Henning:2016lyp}
B.~Henning, X.~Lu and H.~Murayama,
%``One-loop Matching and Running with Covariant Derivative Expansion,''
JHEP \textbf{01}, 123 (2018).
%doi:10.1007/JHEP01(2018)123
%[arXiv:1604.01019].
%85 citations counted in INSPIRE as of 01 Jun 2022

%\cite{Fuentes-Martin:2016uol}
\bibitem{Fuentes-Martin:2016uol}
J.~Fuentes-Martin, J.~Portoles and P.~Ruiz-Femenia,
%``Integrating out heavy particles with functional methods: a simplified framework,''
JHEP \textbf{09}, 156 (2016).
%doi:10.1007/JHEP09(2016)156
%[arXiv:1607.02142].
%82 citations counted in INSPIRE as of 01 Jun 2022

%\cite{Zhang:2016pja}
\bibitem{Zhang:2016pja}
Z.~Zhang,
%``Covariant diagrams for one-loop matching,''
JHEP \textbf{05}, 152 (2017).
%doi:10.1007/JHEP05(2017)152
%[arXiv:1610.00710].
%64 citations counted in INSPIRE as of 01 Jun 2022

%\cite{Beneke:1997zp}
\bibitem{Beneke:1997zp}
M.~Beneke and V.~A.~Smirnov,
%``Asymptotic expansion of Feynman integrals near threshold,''
Nucl. Phys. B \textbf{522}, 321-344 (1998).
%doi:10.1016/S0550-3213(98)00138-2
%[arXiv:hep-ph/9711391].
%761 citations counted in INSPIRE as of 01 Jun 2022

%\cite{Jantzen:2011nz}
\bibitem{Jantzen:2011nz}
B.~Jantzen,
%``Foundation and generalization of the expansion by regions,''
JHEP \textbf{12}, 076 (2011).
%doi:10.1007/JHEP12(2011)076
%[arXiv:1111.2589].
%101 citations counted in INSPIRE as of 01 Jun 2022

%\cite{Cohen:2020qvb}
\bibitem{Cohen:2020qvb}
T.~Cohen, X.~Lu and Z.~Zhang,
%``STrEAMlining EFT Matching,''
SciPost Phys. \textbf{10}, no.5, 098 (2021).
%doi:10.21468/SciPostPhys.10.5.098
%[arXiv:2012.07851].
%17 citations counted in INSPIRE as of 01 Jun 2022

%\cite{Carmona:2021xtq}
\bibitem{Carmona:2021xtq}
A.~Carmona, A.~Lazopoulos, P.~Olgoso and J.~Santiago,
%``Matchmakereft: automated tree-level and one-loop matching,''
[arXiv:2112.10787].
%12 citations counted in INSPIRE as of 01 Jun 2022

%\cite{Criado:2017khh}
\bibitem{Criado:2017khh}
J.~C.~Criado,
%``MatchingTools: a Python library for symbolic effective field theory calculations,''
Comput. Phys. Commun. \textbf{227}, 42-50 (2018).
%doi:10.1016/j.cpc.2018.02.016
%[arXiv:1710.06445].
%36 citations counted in INSPIRE as of 01 Jun 2022

%\cite{DasBakshi:2018vni}
\bibitem{DasBakshi:2018vni}
S.~Das Bakshi, J.~Chakrabortty and S.~K.~Patra,
%``CoDEx: Wilson coefficient calculator connecting SMEFT to UV theory,''
Eur. Phys. J. C \textbf{79}, no.1, 21 (2019).
%doi:10.1140/epjc/s10052-018-6444-2
%[arXiv:1808.04403].
%47 citations counted in INSPIRE as of 01 Jun 2022

%\cite{Gherardi:2020det}
\bibitem{Gherardi:2020det}
V.~Gherardi, D.~Marzocca and E.~Venturini, 
%``Matching scalar leptoquarks to the SMEFT at one loop,''
JHEP \textbf{07}, 225 (2020), [erratum: JHEP \textbf{01}, 006 (2021)].
%doi:10.1007/JHEP07(2020)225
%[arXiv:2003.12525].
%45 citations counted in INSPIRE as of 01 Jun 2022

%\cite{Zhang:2021jdf}
\bibitem{Zhang:2021jdf}
D.~Zhang and S.~Zhou,
%``Complete one-loop matching of the type-I seesaw model onto the Standard Model effective field theory,''
JHEP \textbf{09}, 163 (2021).
%doi:10.1007/JHEP09(2021)163
%[arXiv:2107.12133].
%13 citations counted in INSPIRE as of 01 Jun 2022

\end{thebibliography}
\end{document}